# TOWARD MACHINE LEARNED HIGHLY REDUCED KINETIC MODELS FOR METHANE/AIR COMBUSTION

**Mark Kelly[1], Gilles Bourque[2,3], Stephen Dooley[1]**

[1]Trinity College Dublin, Dublin, Ireland
[2]Siemens Energy Canada Ltd, Montréal, QC H9P 1A5, Canada
[3]McGill University, Montréal, QC H3A 0G4, Canada

## ABSTRACT
*Accurate low dimension chemical kinetic models for methane are an essential component in the design of efficient gas turbine combustors. Kinetic models coupled to computational fluid dynamics (CFD) provide quick and efficient ways to test the effect of operating conditions, fuel composition and combustor design compared to physical experiments. However, detailed chemical kinetic models are too computationally expensive for use in CFD. We propose a novel data orientated three-step methodology to produce compact models that replicate a target set of detailed model properties to a high fidelity. In the first step, a reduced kinetic model is obtained by removing all non-essential species from the detailed model containing 118 species using path flux analysis (PFA). It is then numerically optimised to replicate the detailed model's prediction in two rounds; First, to selected species (OH,H,CO and CH4) profiles in perfectly stirred reactor (PSR) simulations and then re-optimised to the detailed model's prediction of the laminar flame speed. This is implemented by a purposely developed Machine Learned Optimisation of Chemical Kinetics (MLOCK) algorithm. The MLOCK algorithm systematically perturbs all three Arrhenius parameters for selected reactions and assesses the suitability of the new parameters through an objective error function which quantifies the error in the compact model's calculation of the optimisation target. This strategy is demonstrated through the production of a 19 species and a 15 species compact model for methane/air combustion. Both compact models are validated across a range of 0D and 1D calculations across both lean and rich conditions and shows good agreement to the parent detailed mechanism. The 15 species model is shown to outperform the current state-of-art models in both accuracy and range of conditions the model is valid over.*

## INTRODUCTION
To tackle global warming, legislation has been introduced with more stringent targets for greenhouse gas emissions from conventional combustion engines [1]. As a result, gas turbine technology must advance through increasing efficiency, lowering nitrogen oxide production and becoming more flexible to operate on a broader range of fuels. Computational fluid dynamics (CFD) and chemical reactor networks (CRN) are used as rapid development tools to evaluate the effect of combustor design and operating conditions on the performance of the gas turbine engine. These simulations are used as they provide quick and efficient results compared to physical experiments. A key tool needed for a reacting flow simulation is an accurate chemical kinetic model. In such models, the number of transport equations scales with the number of reacting species described by the kinetic model. Detailed chemical kinetic models describe the actual combustion reaction mechanism at fundamental detail and the consequent combustion phenomena to excellent fidelity. However, using this level of detail with CFR or CRN simulations is computationally prohibitive. To allow for rapid application of multi-dimensional simulations, a much smaller chemical kinetic model is highly desired.

Review of the literature shows that there have been two applications of mechanism optimisation. In the first method, the optimisation is performed on a detailed reaction mechanism through uncertainty quantification and minimisation. As Arrhenius parameters in detailed models are derived from physical experiments or theoretical calculations, they have a sizeable uncertainty associated with their value. When they are used in a chemical kinetic model, the model often fails to reproduce, with acceptable accuracy, the results of other physical experiments. This is to be expected as when the uncertainty of each parameter is combined, the resulting uncertainty associated with the model can be quite substantial [2]. It follows therefore that the optimisation of these rate constants, within their uncertainty range, is required to match the model's calculations with experimental measurements within a desired range of conditions that the mechanism will be valid over. Most works of this type apply variants of the *solution mapping* method originated by Frenklach and Miller [3, 4]. This involves the generation of *response surfaces* for each optimisation target. A model's prediction of the optimisation target is then analytically obtained from the mathematical relation of the response surface instead of performing an actual simulation, which is more costly. This dramatically reduces the computational time, enabling the use of laminar flame speeds as optimisation targets. This approach to parameter estimation has been implemented and modified by various authors, notably by Wang and Sheen in the development of their MUM-PCE method [5-8] and by Turányi [9] who made use of wider and more comprehensive range of direct and indirect experimental measurements as optimisation targets.

The second method of mechanism optimisation, and that of more interest to this work, is the optimisation of reduced model's parameters to produce highly accurate compact kinetic models. Compact models do not contain the necessary number of species in the reaction network to accurately describe the combustion process, i.e. the chemical mechanism is insufficiently described. Consequently, models that use a species reaction network with this limited number of nodes (species) have a poor replication of the target combustion either computed by a detailed model or known from experiment.

Many reduced models in literature make use of analytical relations to remove species from the reaction network(e.g. Quasi-Steady State assumption) , use correction factors  that are a function of the specific operating conditions and/or use non-standard reaction rate descriptions to minimise the number of species contained in the model. Although there are benefits to these techniques, their use limits the deployment of the model. Modification of the numerical solver is a non-trivial exercise particularity when using standard industry packages such as Cantera [10], ChemKin [11], STAR-CCM+ [12], ANSYS Fluent [13] etc.

The FAIR [14] principles of data science require the models to be interoperable, meaning they can be readily used in current software infrastructure. Therefore, the aim of this work is to produce compact models that can be easily incorporated into current software infrastructure without modification (*"plug and play"*).

Table 1 shows the current state-of-art methane compact models that meet these requirements. State-of-art is defined regarding accuracy in terms of performance and number of species in the model.

**Table 1.** State-of-art compact methane/air combustion models that require the use of no additional numerical solver subroutines for their application.

| Authors | Size | Optimisation Technique | Parent Mechanism |
|---|---|---|---|
| Lytras et al. [15] | 14 species<br>23 reactions | Not described | USC Mech II [16] |
| Leylegian [17] | 15 species<br>51 reactions | Steepest Descent Method with Solution Mapping | USC Mech II [16] |
| Luca et al. [18] | 16 species<br>72 reactions | N/A | GRI 3.0 [19] |
| This Work | 15 Species 55 Reactions | Machine Learned Optimisation of Chemical Kinetics | NUIG18_17_C3 [20] |

In this study we show how a compact model can be obtained first by a careful reduction of the number of species followed by an optimisation performed in a semi-autonomous manner. The compact model will be demonstrated for methane combustion over a wide range of operating conditions.

## METHODOLOGY

Compact models comprise four components:

1) A virtual reaction network in place of the reaction mechanism. This serves the purpose of providing sufficient degrees of freedom to the compact model such that the reaction flux may be sufficiently regulated between the network nodes (species), to replicate a defined set of combustion properties. The more complex the combustion properties set, over the wider range of conditions, the more complex a reaction network is needed.
2) Effective reaction rate constants, analogous to the elementary reaction rate constants, but of no physical reality. These describe the connection between every network node.
3) Thermodynamic data for each species. These are used to calculate thermodynamic properties for each species and equilibrium constants for each reaction.
4) Transport property data for each species describing species transport. These are used in the solving of the transport/conservation equations.

In our approach, components 1 and 2 are manipulated in a semi-autonomous manner whereas components 3 and 4 are kept at the fundamentally derived values used in the detailed model.

Our state of art in this regard is given by models produced by the *path flux analysis (PFA)* [21] method, which is a physics-based algorithm that simplifies the chemical mechanism by the removal of species that are relatively less important in the grand scheme. However, as species are removed, the fidelity of the calculations of the reduced model gradually deteriorate until a critical point of necessary detail is reached, after which further removal of detail renders the model completely inaccurate even for the most basic phenomenon. This is due to the over-reduction of the mechanism, causing the authentic reaction flux to be severely compromised to the reality. For methane combustion, this threshold occurs at approximately 25-30 species (Figure 11), depending on the range of conditions and phenomena the model is required to be valid over. However, compact models with less than 20 species are required for CFD to be computationally affordable.

In this study, to fix this, effective reaction rate constants of the retained reactions are altered to divert flux through the system such that the model can accurately replicate the parent detailed model's chemistry over a range of operating conditions and reactor geometries. This method implicitly assumes that the *"flux map"* computed by the detailed model chosen, is an accurate representation of the reality. It can be expected that this is the case for well-studied fuels such as methane. However, if not well-known fuels are the subject, the method can be adjusted to use experimentally determined information, or a combination of modelling data with experimental data.

For compact models the original parameters of the detailed model are no longer a fair approximation of the effective reaction flux through a set of species. The task of the optimisation is to adjust these effective reaction rate constant parameters to embody the effective reaction flux through the more limited set of species in the overly reduced model, such that the same behaviors embodied in the detailed model are produced.

Genetic algorithms [22] are the most used method to optimise compact models [23-27]. Although the exact procedure for each genetic algorithm varies, the general approach is similar as follows;

i. A series of kinetic models are initially generated with different values of the parameters to be optimised.
ii. From this set, the best performing models are selected as the *fittest*. Another set of models are then generated based on the parameters of the previously found fittest models.
iii. These are evaluated, and the best performing models selected as the new fittest models.

At each step, random perturbations of the parameters occur to prevent the algorithm from falling into a local minimum. These are called *mutations*. This process repeats until the error on the fittest model is below a predefined threshold. The models are typically evaluated on their ability to replicate flame profiles and ignition delay times of the parent detailed mechanism. The method used in this work shows some similarity to these algorithms.

### Methane Detailed Mechanism Reduction and Optimisation

Path flux analysis (PFA) was applied to reduce the NUIG18_17_C3 detailed model comprising 118 species to its lowest threshold of size, producing one model with a reaction network comprising 19 species and one with a reaction network comprising 15 species.

To perform a PFA reduction, a numerical database of time dependent chemical species concentration profiles in a homogenous constant volume reactor up until the point of ignition is required. This is performed using the SENKIN [29] programme which defines the point of ignition as the point at which the temperature first exceeds 400 K above the initial temperature. The database is populated with simulations using methane/air mixtures at 1, 10, 20, 30, and 40 atmospheres (atm), 900 – 2000 K and at equivalence ratios 0.5, 1 and 1.5. To examine the extinction phenomenon, a second numerical database populated at conditions of 1, 10, 20, 30 and 40 atm, 300 - 1500 K and equivalence ratios 0.5, 1 and 1.5 is constructed using perfectly stirred reactor (PSR) simulations.

The resulting 19 and 15 species models (reaction network + effective reaction rate constants + thermodynamic and transport properties) perform poorly against these validation targets as the reaction rate constants are not a sufficient descriptor of the reaction flux through this reaction network. Optimisation was thus performed to improve their fidelity by altering the flow of flux through the system such that it better resembled that of the detailed model over the range of conditions of interest.

To do this, an in-house *Machine Learned Optimisation of Chemical Kinetics (MLOCK)* algorithm has been developed at TCD. This paper reports only the work of *MLOCK_ver1* which optimises the effective reaction rate constants of all reactions in the reaction network that show influence on the computation of several optimisation targets, by perturbation of each of the Arrhenius rate constant parameters. All combustion simulations were performed using the Cantera package implemented through python.

## Machine Learned Optimisation of Chemical Kinetics (MLOCK)

The MLOCK algorithm is a two-phase optimisation procedure that optimises compact models to replicate the detailed model's behaviours. The algorithm produces and evaluates new models through the generation of new $A$, $n$ and $E$ Arrhenius rate constant parameters for selected reactions. An optimisation target, in this case specified calculations of the detailed model, is selected to form a series of Objective Error Functions *(OEF),* which evaluates the suitability of the new set of effective reaction rate constants. Figure 1 sumarises the overall optimisation method as a series of steps, split across two phases.

Each phase is composed of several *scans*. These scans search an area of the parameter bound space (effective rate constant magnitude) to assess the suitability of that region in producing accurate models. Initially, a large coarse grid-like scan is performed to generate an approximate representation of the parameter bound space, searching an effective reaction rate constant range of approximately 15 orders of magnitude. Approximately 80,000 discrete models are produced, and their performance evaluated. From this, areas that produce models showing low OEFs are identified. A second scan of (another) 80,000 models is then performed with the search algorithm constrained within narrower bound spaces, centered on regions showing low OEF, identified in the initial scan. This thereby increases the resolution in these regions. From these two initial scans, the best performing model is identified and selected as a *genetic seed*. This procedure is analogous to the *fittest models* approach used in genetic algorithms. Subsequent scans are centered on this genetic seed and focuses the search algorithm in a region of the bound space that can produce a larger number of accurate models and models of improved accuracy.

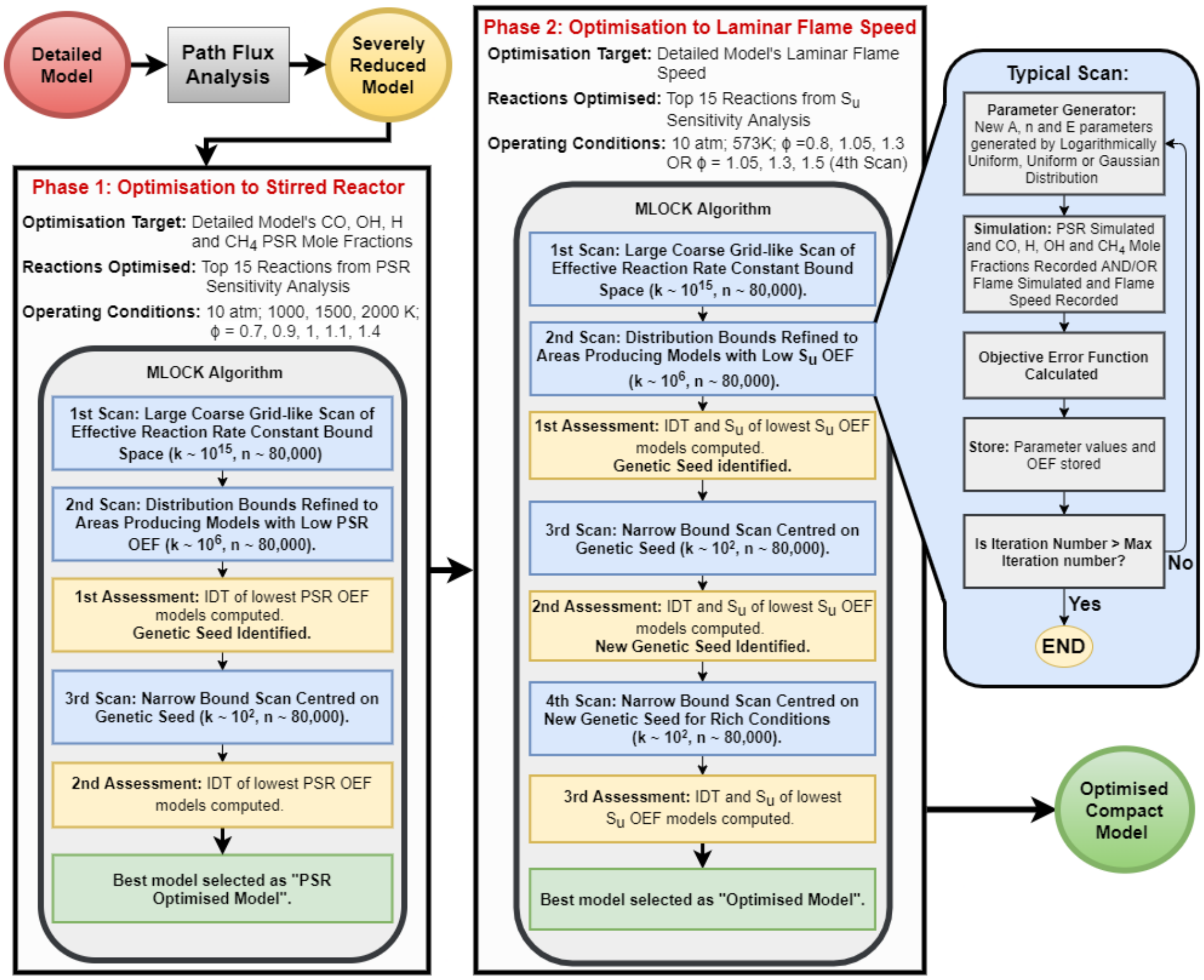

**Figure 1: The two-phased *MLOCK_ver1* optimisation procedure. The range of effective reaction rate constant values searched, and the number of models produced at each step is noted.**

## MLOCK Phase 1

The first phase of the MLOCK_ver1 is optimisation of the compact model to detailed model perfectly stirred reactor (PSR) calculations. A PSR is used as it assumes a perfectly homogeneous mixture, removing any mixing limitations and ensuring the system is kinetically driven. Therefore, if the compact model can replicate the detailed model's PSR response then it can correctly predict the reaction kinetics of the system and other kinetically driven combustion properties. As the performance of a model is evaluated by the OEF, the selection of properties to evaluate and the mathematical form of the OEF is of key importance.

The mole fraction profiles of OH, H, CO and $CH_4$ were selected as the optimisation targets to comprise the OEF for the first optimisation phase. These species were selected as they are either essential intermediates, contributors to heat release or are indicators for fuel decomposition. The accurate replication of these key species encourages an accurate replication of the relevant kinetics of the detailed model. A PSR simulation is performed using each model and these mole fractions recorded. This is performed at 10 atmospheres (atm); 1000, 1500 and 2000 Kelvin (K) and equivalence ratios 0.7, 0.9, 1, 1.1, 1.4 to capture lean, stoichiometric and rich chemistry at gas turbine relevant pressure and temperature. The OEF is then calculated and the values stored.

As we are mostly interested in the calculation of the species concentrations when the system is kinetically driven, i.e. at times close to ignition, an OEF is needed that prioritises the error in concentrations in this region. This is shown in Equation 1. The first term in this OEF is designed such that errors in concentrations close to the maximum concentrations are weighted more heavily than those far from the peak concentration. This term was used by Turányi et al. to evaluate the error in species concentrations during model reduction [30]. As the intermediates OH, H and CO experience a peak in concentration at time of ignition, they are included in this term. However, as fuel, $CH_4$ always has a maximum concentration at the start of the simulation (t=0) this term does not provide a suitable evaluation of its error. Therefore, the error in $CH_4$ concentration is evaluated by a simple squared error.

$$PSR\ OEF = \frac{1}{N_J N_k}\sum_k \sum_j \left\{ \left[ \sum_i w_i \left( 2\frac{\left| X_{i,k}^{red}(t_j) - X_{i,k}^{det}(t_j) \right|}{X_{i,k}^{det}(t_j) + max\left(X_{i,k}^{det}\right)} \right) \right] + w_{fuel}\left( X_{fuel,k}^{red}(t_j) - X_{fuel,k}^{det}(t_j) \right)^2 \right\} \quad (1)$$

Where, $X_{i,k}$ refers to the mole fraction of species $i$ (OH, H, CO) at the $k$'$^{th}$ set of temperature, pressure and equivalence ratio conditions. $N_k$ and $N_J$ respectively refer to the total number of operating conditions and time points that are included in the evaluation. $N_k$ and $N_J$ equal fifteen and three thousand respectively in this work. The weighting term, $w$ is used to assign a greater importance to the error of certain species if desired. However, in the current study this is kept at unity. Figure 2 shows the suitability of this OEF as an indicator for ignition delay time, an important combustion property that is commonly used to evaluate the performance of a model in reproducing the kinetics of a detailed model. This shows the importance of the accurate replication of these PSR profiles.

## MLOCK Phase 2

Following optimisation to the detailed model's PSR response, the best performing model is identified and subsequently optimised to the detailed model's calculations of the laminar flame speed. Following Figure 1, Phase 2 is similar to Phase 1 up until the *2^nd^ Assessment* step. However, subsequent to this second assessment step, a fourth scan is performed with a new genetic seed found from the third scan.

In the first three scans the laminar flame speed was evaluated at 10 atm, an unburned gas temperature of 573 K and (fuel/air) equivalence ratios 0.8, 1.05 and 1.3 to account for both lean and rich behaviours. It was found that numerous models were produced that could accurately replicate the laminar flame speed at lean conditions, however their performance at rich conditions varied. This necessitated a fourth scan that optimised the model to rich conditions exclusively, i.e. at equivalence ratios 1.05, 1.3 and 1.5.

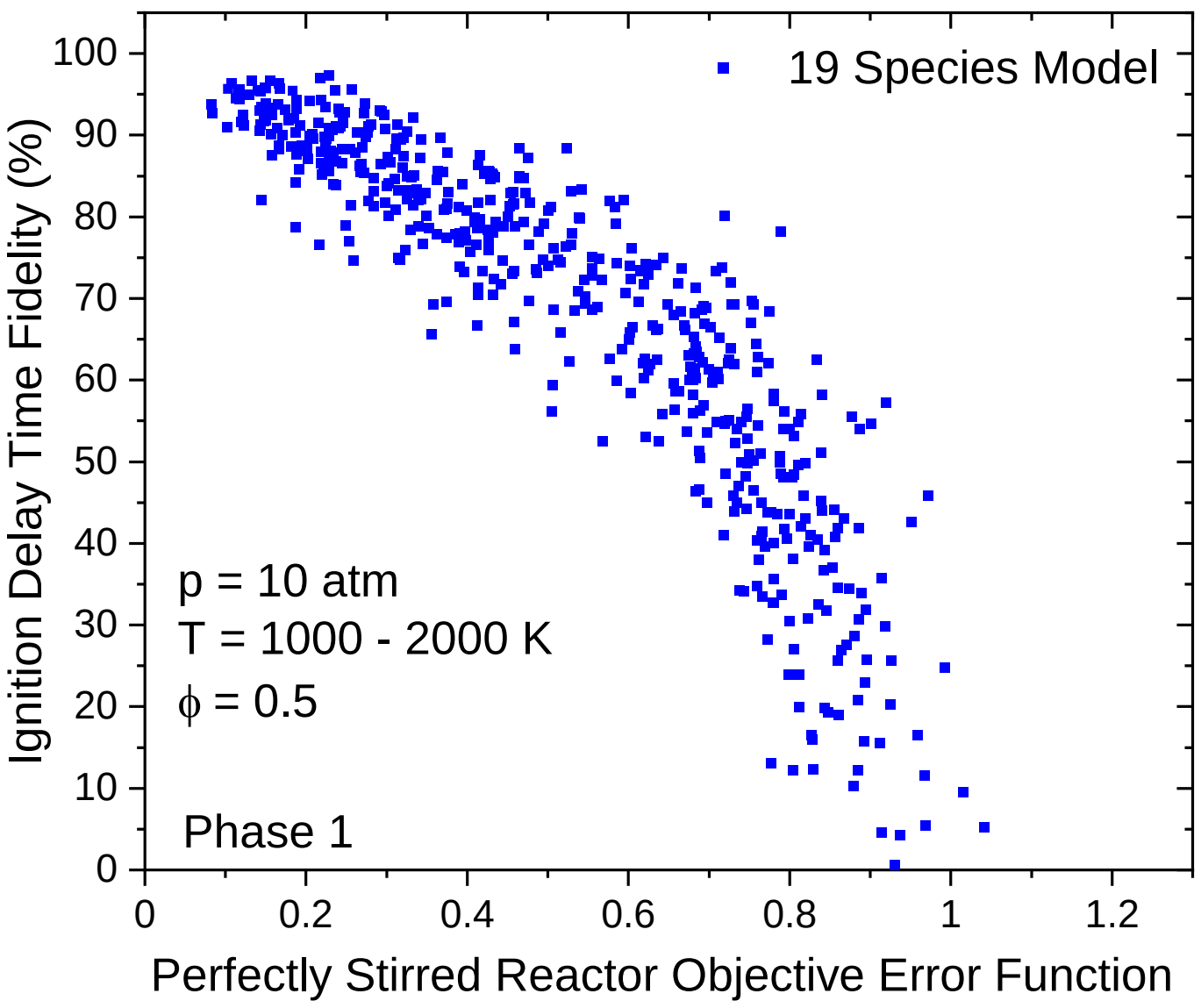

**Figure 2: Ignition delay time fidelity versus perfectly stirred reactor objective error function (PSR OEF) for one thousand iterations of the 19 species methane reaction networks as produced by MLOCK_ver1 Phase 1.**

The objective error function used in Phase 2, the $S_u$ OEF, is shown in Equation 2 below. As the error in each flame speed measurement is of equal importance, a simple relative error is used.

$$S_u\ OEF = \frac{1}{N_P}\sum_p \left( \frac{\left| S_{u,p}^{\ red} - S_{u,p}^{\ det} \right|}{S_{u,p}^{\ det}} \right) \quad (2)$$

Where $S_{u,k}$ is the laminar flame speed at the p'$^{th}$ condition set of temperature, pressure and equivalence ratio, and $N_P$ is the number of conditions, which in the current study equals three. The best performing model resulting from this second phase of optimisation is identified as the optimised model.

## RESULTS AND DISCUSSION

The optimisation methodology proposed is demonstrated by producing compact models constructed on reaction networks comprising 19 species and 15 species for methane/air combustion. These optimised models are referred to as 19sOp and 15sOp hereinafter. The models can be found at the TCD Energy Research webpage [31]. Both models have been exposed to only the simulations, and at conditions, presented in Figure 1. The validity of each model therefore is tested at a range of zero-dimensional (0D) and one-dimensional (1D) configurations that expand outside of this range.

To quantify the performance of each model, a Fidelity Index is defined as a simple sum of the relative errors of each point in the data set, as shown in Equation 3.

$$Fidelity\ (\%) = \frac{100}{N}\sum_k \left(1 - \frac{\left|X_k^{compact} - X_k^{detailed}\right|}{X_k^{detailed}}\right) \tag{3}$$

where $X_k$ is the IDT or laminar flame speed at the n'$^{th}$ set of pressure, temperature and equivalence ratio conditions and $N$ is the number of conditions included.

### Zero-Dimensional Simulations

Simulations that use a 0D reactor, assume the mixture to be homogenous, thereby eliminating the effect of species transport and diffusion. These simulations are used as validation tests to assess the ability of the compact models in replicating the reaction kinetics and thermochemical behaviours of the detailed model. These validation checks include; PSR profiles of important species, evolved PSR temperatures, adiabatic flame temperatures and ignition delay times.

PSR simulations are used as development tools by gas turbine engineers. The primary zone of a combustor can be approximated and simulated as many interconnected PSRs in chemical reactor network (CRN) techniques [32, 33]. Therefore, it is important for the compact model to accurately reproduce PSR profiles, particularly the evolved temperature and species concentrations. The optimisation process greatly improved both model's replication of the detailed model's PSR response, as expected. PSR simulations were performed across the range of conditions as shown in Table 2. For the sake of brevity only selected conditions are presented here to serve as representative of the typical performance of the models. Figure 3 shows the temperature of a PSR as a function of residence time for a stoichiometric methane/air mixture at 10 atm and 1500 K using detailed and both compact models. The 19sOp replicates the detailed model's temperature perfectly with 15sOp showing a slight discrepancy in predicting the extinction residence time. 15sOp predicts that the minimum residence time to ensure ignition at this set of conditions is 0.1 ms, whereas the detailed model predicts this to be 0.14 ms. Figure 4 shows the burned gas temperatures of a PSR using the detailed and compact models as a function of equivalence ratio. The PSR here has a residence time of 1 ms and an inlet temperature of 1500 K. Both compact models replicate the temperatures excellently with the average deviation between the detailed model and the 15sOp model of 0.05 percent.

**Table 2. Range of validation conditions.**

| Validation Property | Pressure (atm) | Equivalence Ratio | Temperature (K) |
|---|---|---|---|
| Ignition Delay Time (IDT) | 1 – 40 | 0.5, 1, 1.5 | 1100 - 2000 |
| Laminar Flame Speed ($S_u$) | 1 – 40 | 0.4 – 1.5 | 473 – 673 |
| Perfectly Stirred Reactor Profiles ($OH$, $H$, $CO$, $CH_4$) | 10 | 0.7 - 1.4 | 1000 - 2000 |

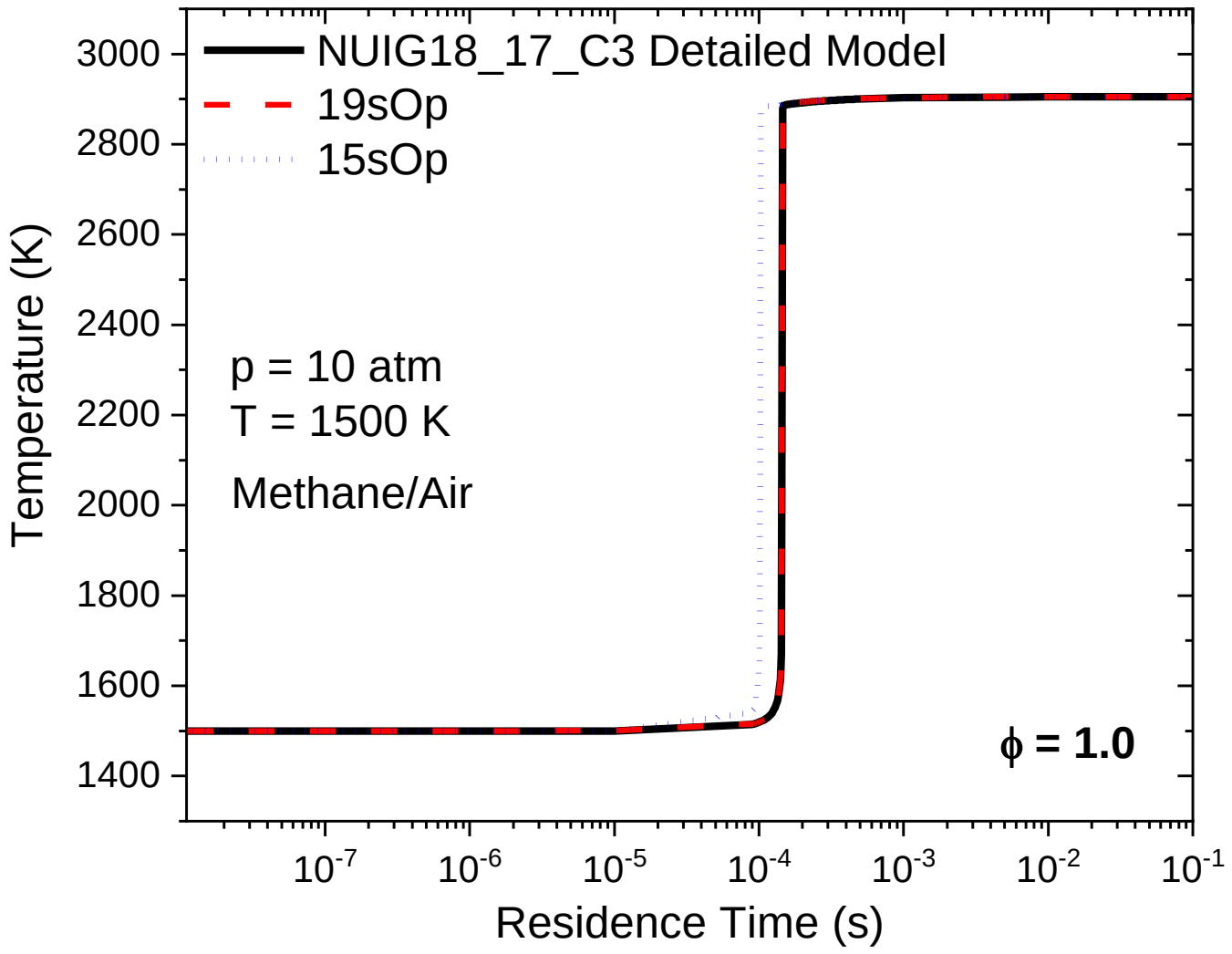

**Figure 3: PSR temperature as a function of residence time at 1500 K, 10 atm for stoichiometric methane/air as calculated by the detailed (solid), 19sOp (dash) and 15sOp (dot) models.**

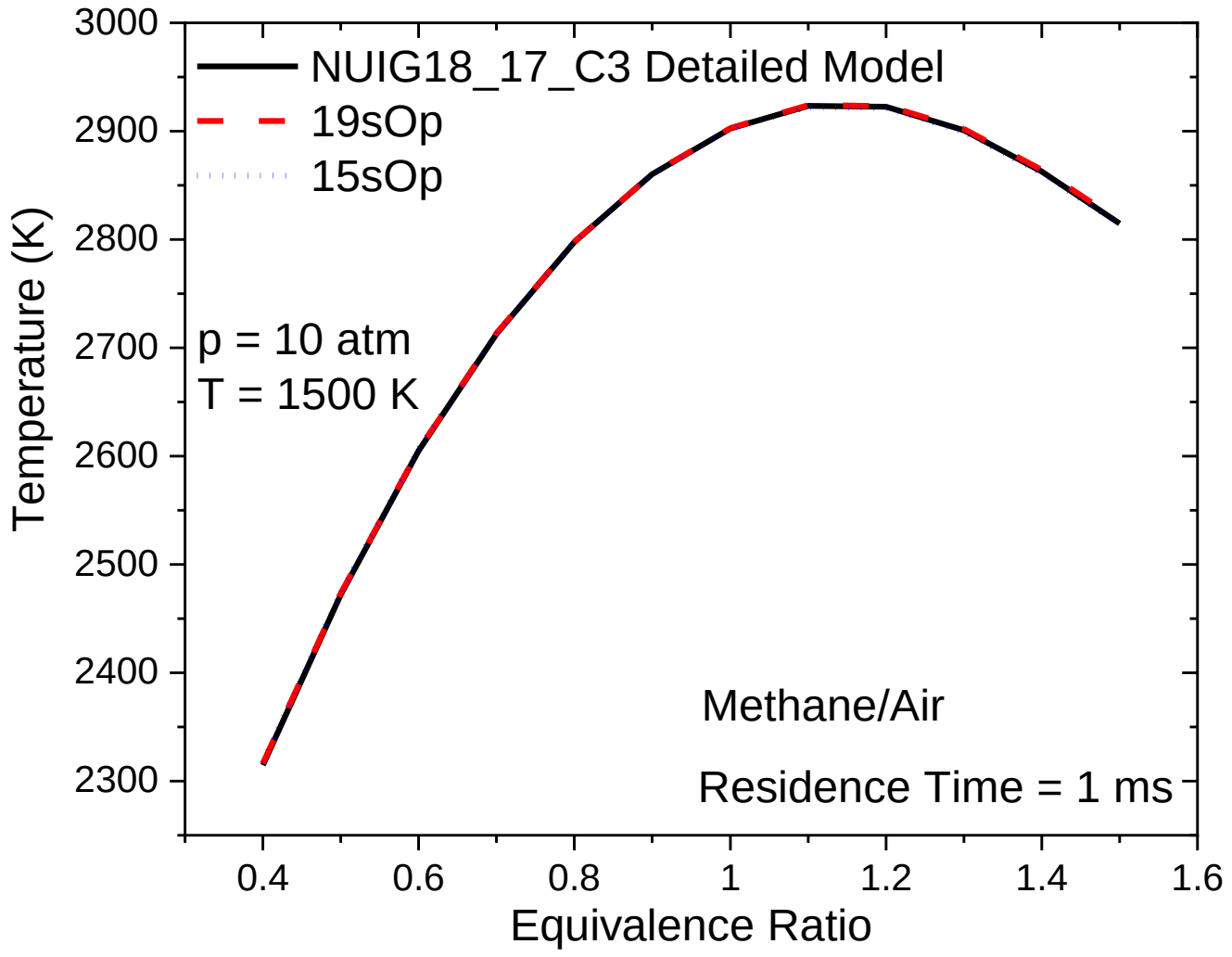

**Figure 4: PSR temperature with residence time of 1 ms and an initial temperature of 1500 K as a function of methane/air equivalence ratio as predicted by the detailed (solid), 19sOp (dash) and 15sOp (dot) models.**

The 19sOp has an excellent replication of PSR mole fraction profiles across all conditions. Although 15sOp tends to predict well the peak species mole fractions, there is a time resolution error that varies as a function of operating condition. The performance of 15sOp is notably poorer than the 19sOp model for low temperature rich mixtures. This is likely due to the removal of ethane ($C_2H_6$) and methoxy ($CH_3O$) from the system which are both important in capturing methyl recombination and decomposition chemistry respectively at low temperature and rich methane mixtures.

Figure 5 shows the PSR profiles of key species at a representative set of operating conditions to illustrate the typical performance of each model. It can be seen that 15sOp has a time resolution error that scales with ignition delay time error, however its calculation of the magnitudes of key species is relatively accurate, as shown in Figure 6.

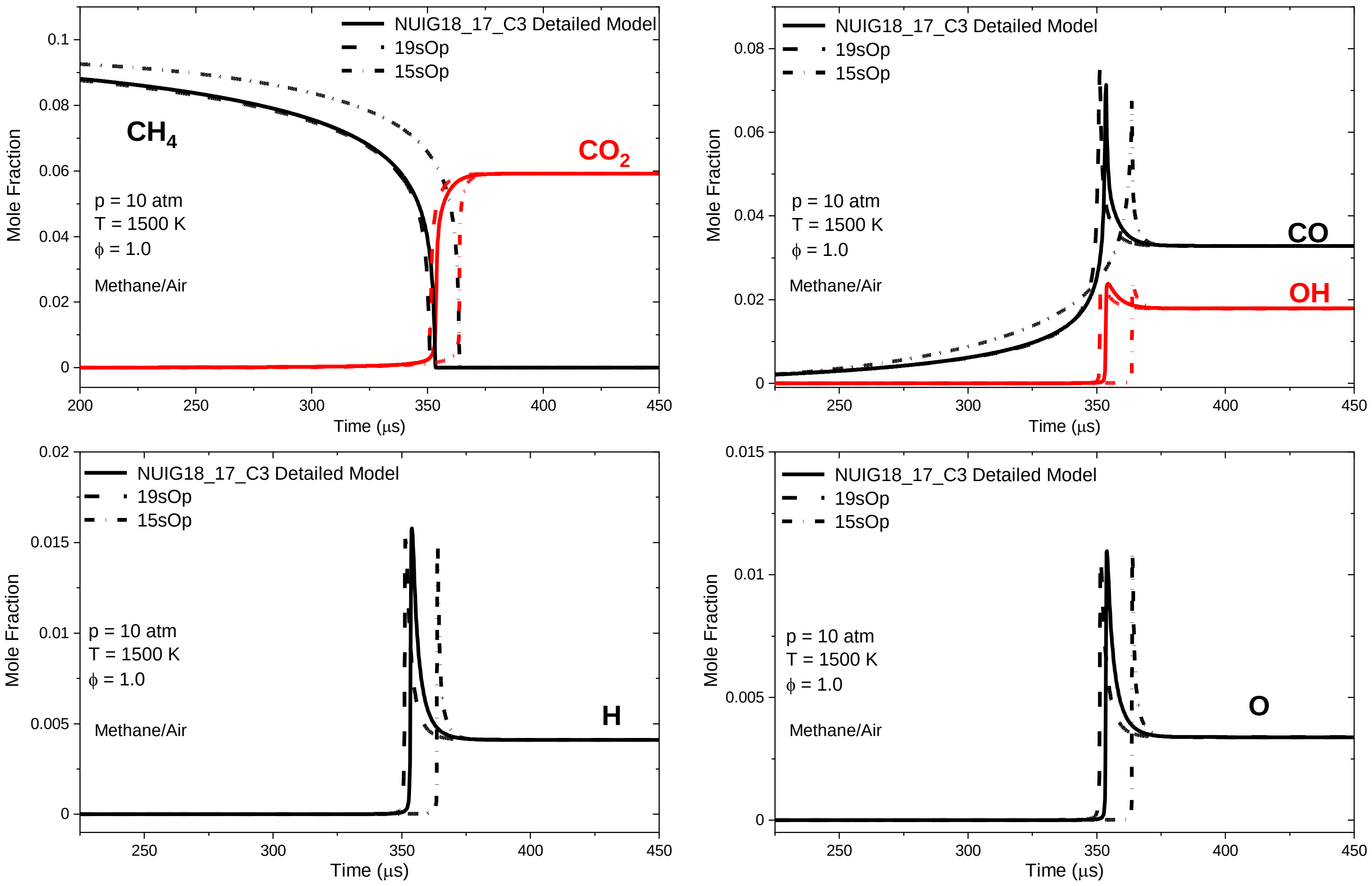

**Figure 5: Perfectly stirred reactor mole fraction profiles of important species at 10 atm, 1500 K for stoichiometric methane/air using the detailed (solid), 19sOp (dash) and 15sOp (dot) models.**

Ignition delay time (IDT) is a kinetically driven phenomenon that is used to examine the reactivity of the fuel mixture at a set of conditions. This is commonly used as a validation test to analyse the performance of a chemical kinetic model. The IDT in this case was defined as the time point at which the derivative of temperature with respect to time was a maximum. This was performed using the detailed and both compact models at the range of conditions shown in Table 2. The performance of the models is shown in Figure 7. 19sOp predicts ignition delay times within 95 percent of the detailed model across this range of conditions. 15sOp also has a good fidelity (75 percent), with the performance worsening at low temperatures, for the reasons described earlier, reducing the overall fidelity. Although the models were only optimised at 10 atm, the models can predict the ignition delay times to a similar degree of accuracy up to 40 atm, however the accuracy decreases when reduced to atmospheric pressure. To increase the performance at lower pressures, the PSR OEF could be altered to include atmospheric conditions as optimisation targets.

## One-Dimensional Simulations

In 1D simulations, the effect of species transport is accounted for. This added level of complexity significantly increases the computational time to solve. The laminar flame speed is an intrinsic characteristic of the fuel. As such, it plays a key role in understanding both the kinetics and diffusivity of the fuel mixture and how the two phenomena interact. Hence, it is widely used as a development tool and as a performance test for chemical kinetic models.

The premixed laminar flame speed was computed at 1, 10, 20, 30 and 40 atm, unburned gas temperatures 473, 573 and 673 K and equivalence ratios 0.4 to 1.5 at 0.1 intervals. Both models have good replication of the laminar flame speed predictions of the detailed model as shown in Figure 8. 19sOp replicates the detailed model to over 95% at the set of operating conditions shown. 15sOp replicates the detailed model to over 92%, with an over prediction by approximately 10% at slightly rich mixtures. Although 15sOp was only optimised to laminar flame speeds at

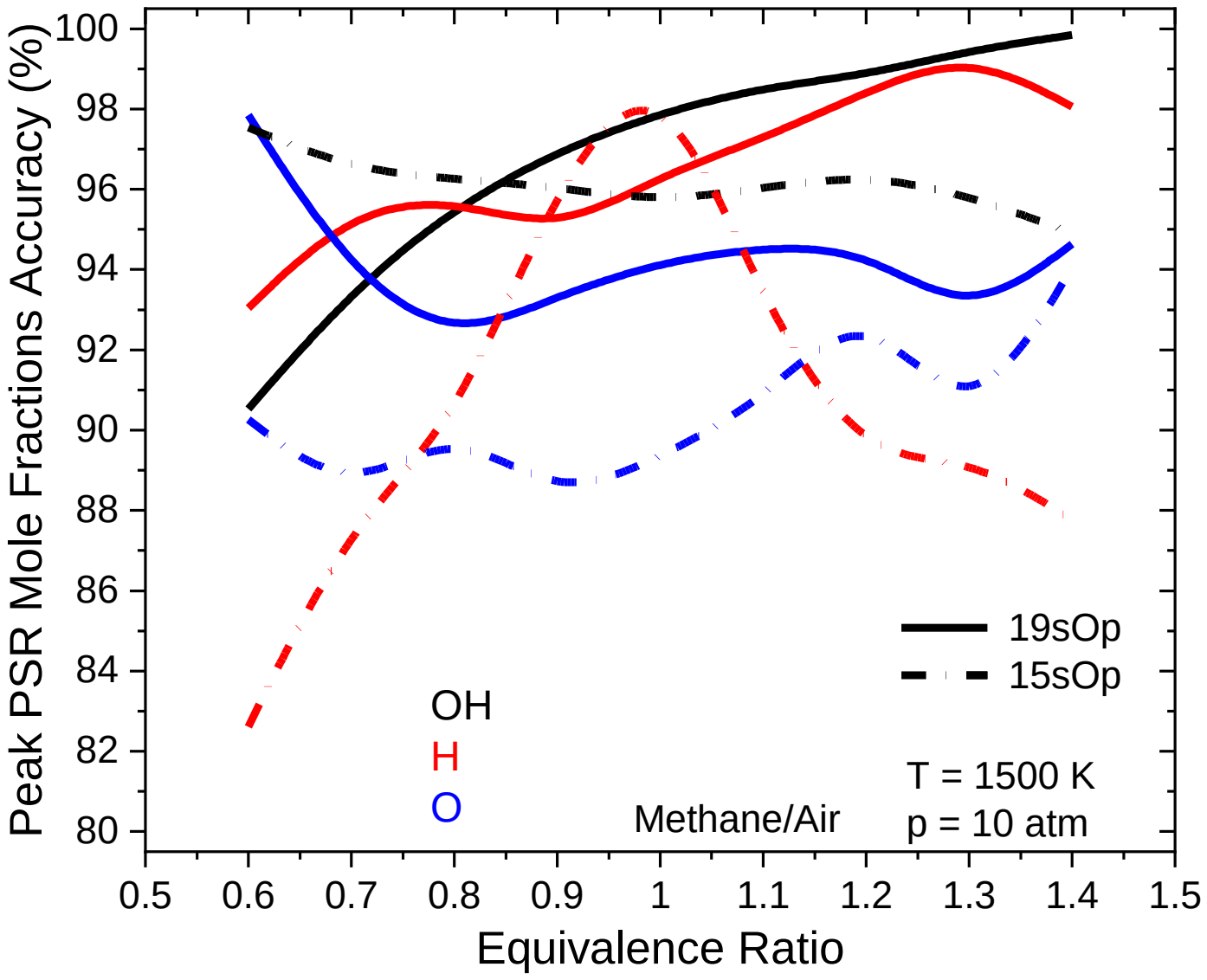

**Figure 6: Performance of 19sOp (solid) and 15sOp (dot) in predicting maximum mole fraction of important intermediate species in a PSR for methane/air at 10 atm, 1500 K and ϕ = 1.**

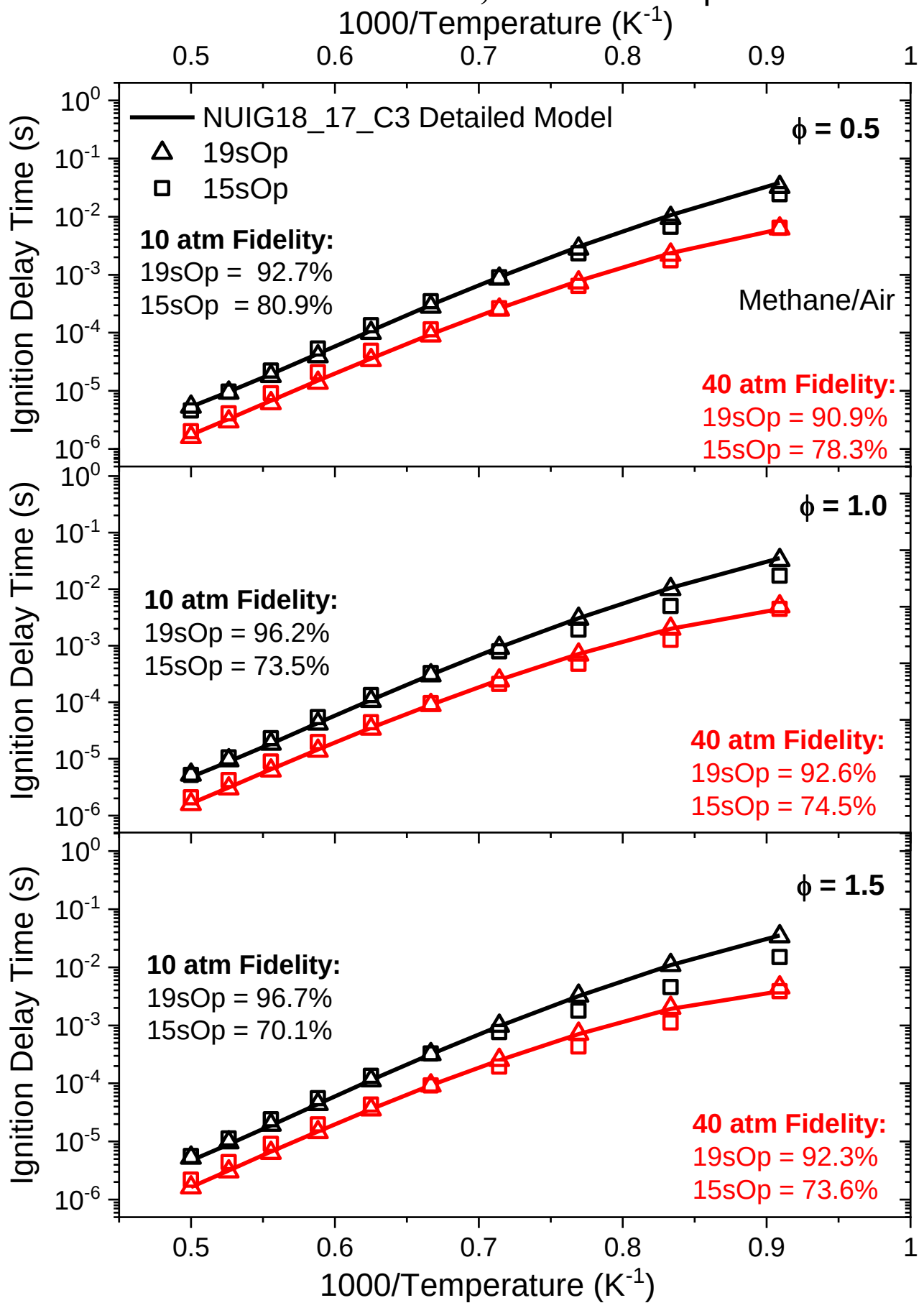

**Figure 7: Ignition delay times of the detailed (line), 19sOp (triangle) and 15sOp (square) models at 10 and 40 atm for lean, stoichiometric, and rich methane/air mixtures.**

10 atm and with an unburned gas temperature of 573K, it can predict flame speeds at higher and lower temperatures to a similar accuracy. 19sOp can predict flames speeds up to 40 atm with a similarly good accuracy as at 10 atm, however, 15sOp becomes less accurate as the pressure is changed from 10 atm. Figure 9 shows the laminar flame speed of the detailed, 19sOp and 15sOp models at lean and stoichiometric methane/air mixtures as a function of pressure. Both models' performance decreases as the pressure is reduced to atmospheric. To increase the accuracy of the models at these pressures, an atmospheric condition must be included as an optimisation target.

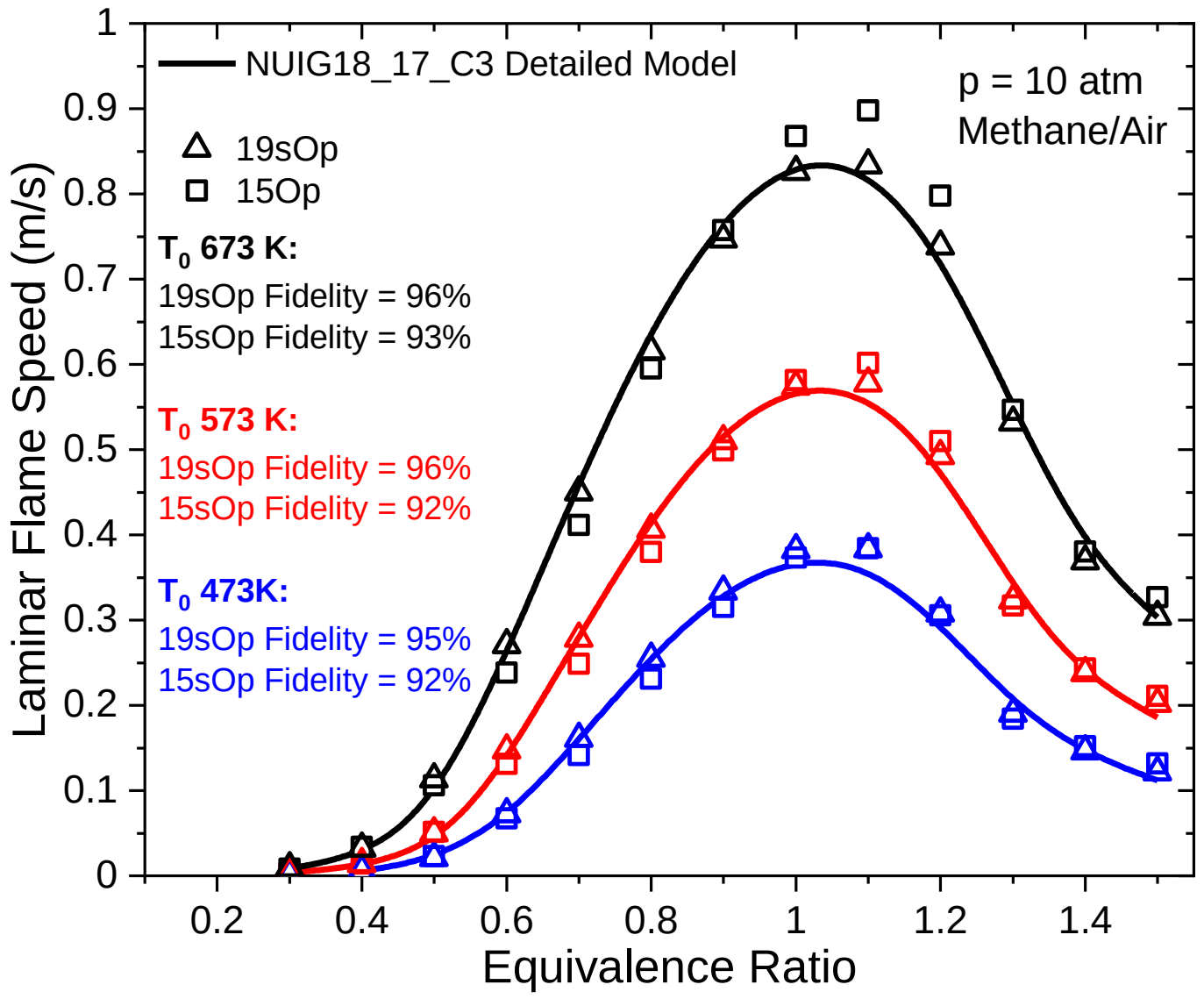

**Figure 8: Laminar flame speed as a function of equivalence ratio for methane/air using the detailed (line), 19sOp (triangle) and 15sOp (square) models at 10 atm for varying unburned gas temperatures.**

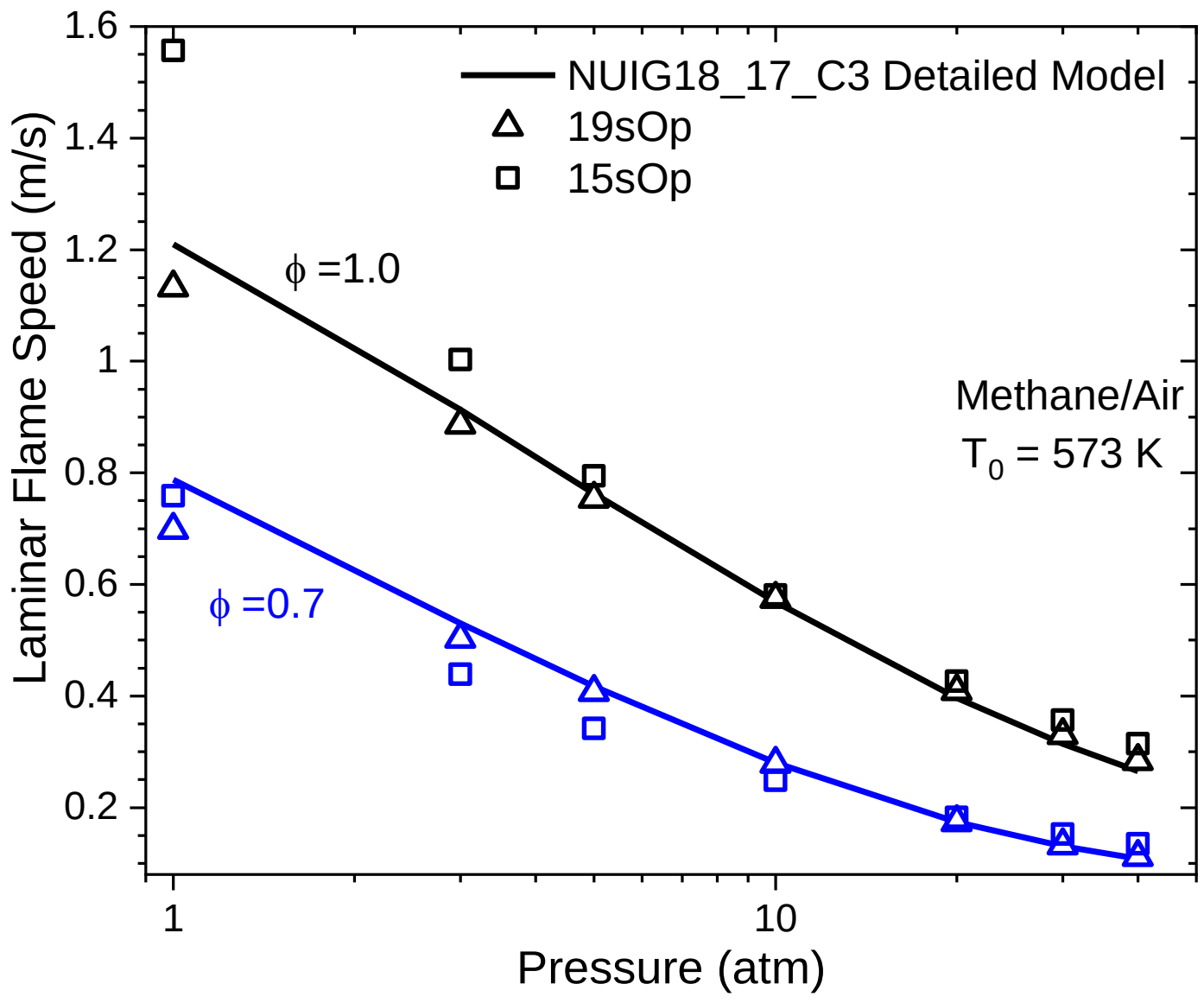

**Figure 9: Laminar flame speed as a function of pressure for stoichiometric and lean methane/air mixtures using the detailed (line), 19sOp (triangle) and 15sOp (square) models.**

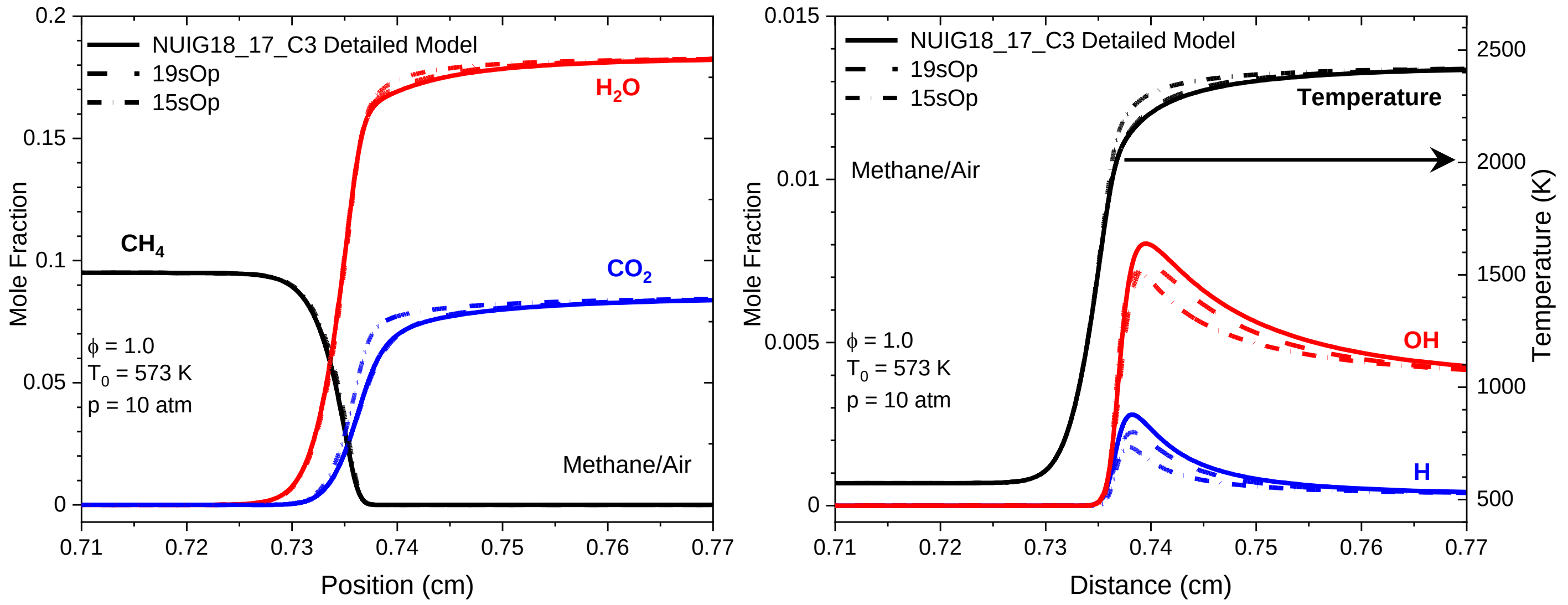

**Figure 10: Flame structure of a stoichiometric methane/air laminar premixed flame at 10 atm with an unburned gas temperature of 573 K using the detailed (solid), 19sOp (dash) and 15sOp (dot) models.**

The mole fractions of several important species in a premixed laminar flame calculated by the detailed and compact models are shown in Figure 10. 19sOp has a good replication of the detailed model's flame structure across a wide range of conditions. The model predicts well the peak concentrations of many important species. 15sOp too has a good replication of many major species however it tends to under predict the peak concentration. Both models also replicate the temperature in the flame well.

The computational time to solve a 1D flame using 19sOp and 15sOp is 1.3% and 0.9% of the time it takes the detailed model. This is a speed up factor of 79 and 106 respectively.

In the production of the 19 and 15 species models by PFA, the detailed model was reduced with increasingly higher thresholds, producing a series of reduced models of various sizes ranging from 42 to 15 species. The ignition delay time at lean conditions and laminar flame speed were calculated using each of these models and their fidelities calculated. The results are shown in Figure 11. It is shown that the performance of the model begins to worsen considerably when the number of species drops below 30. This is evidence of the threshold of necessary detail described earlier. The figure shows that the ignition delay time fidelity of the 15 and 19 species models are increased through the optimisation procedure from 60% to 86% for the 15 species model and 72% to 95% for the 19 species model over the range of conditions noted. The laminar flame speed fidelities are similarly improved from 65% to 83% for the 15 species model and 77% to 96% for the 19 species model.

## Comparison to State-of-Art

Table 3 compares 19sOp and 15sOp with models of broadly equivalent sizes available in literature across a wide range of gas turbine relevant conditions as defined by Table 2, all for methane/air. The 21 species DRM19 and 24 species DRM22 [34] models are included for familiarity as they are well-known and well performing historical benchmarks. As shown in Table 2, 15sOp outperforms the state-of-art models in terms of both 0D and 1D combustion phenomena. It is important to note that only the DRM models and those of the current work are created for the purpose of describing a broad range of conditions. The models of Luca et al. and Lytras et al. were constructed to be accurate for a flame at atmospheric pressure and so as expected these models perform badly versus 0D validation targets and at elevated pressures.

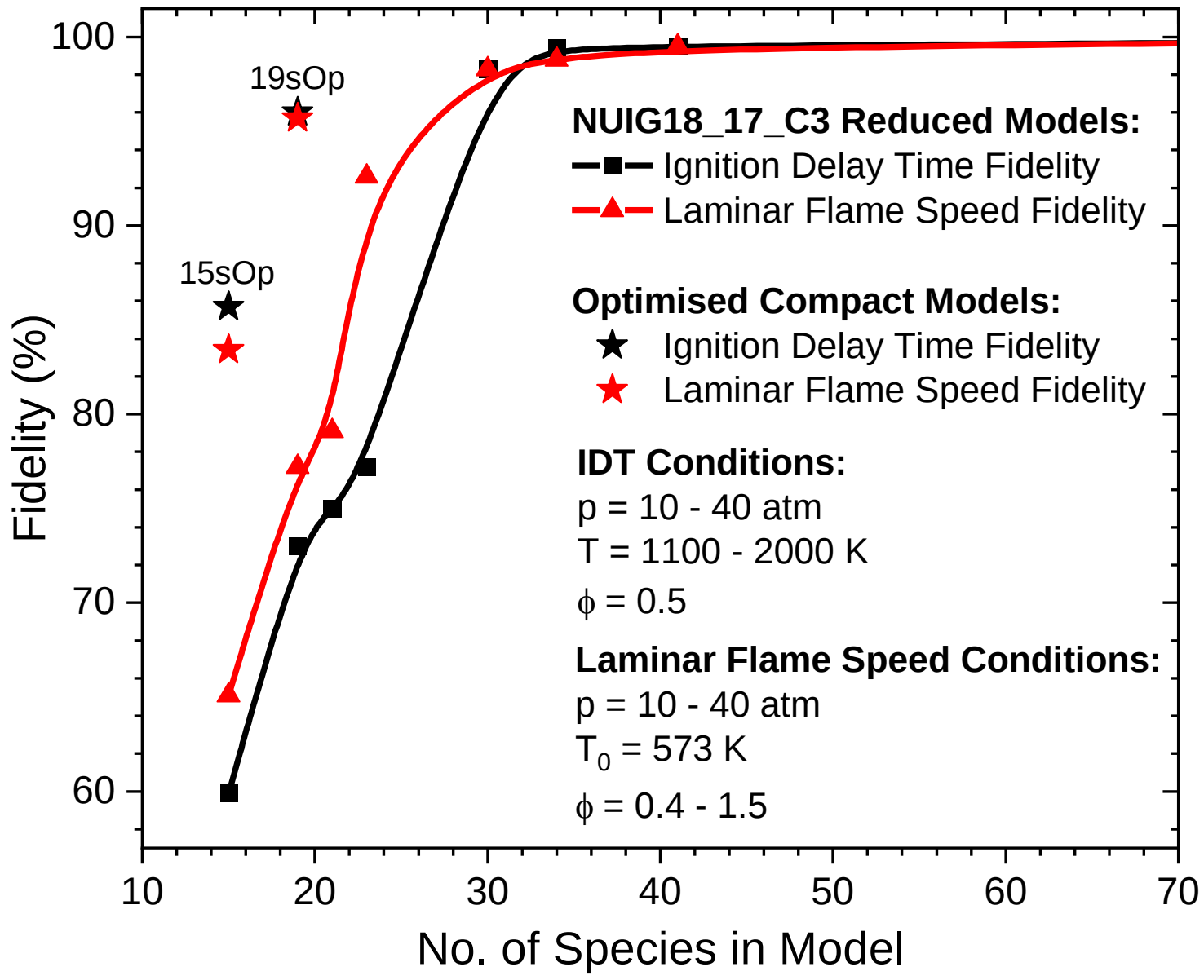

**Figure 11: The effect of the number of species in the model on ignition delay time (black) and flame speed (red) fidelity. Fidelity of compact models produced in this work shown (star).**

**Table 3. Performance of state-of-art compact methane/air combustion models.**

| Author | No. of Species | IDT Fidelity | $S_u$ Fidelity | PSR Fidelity | Parent Mechanism |
|---|---|---|---|---|---|
| Lytras | 14 | 18.6% | 65.4% | 32.2% | USC Mech II |
| Leylegian | 15 | 39.3% | 83.6% | 40.8% | USC Mech II |
| Luca | 16 | 0% | 65.4% | 5.5% | GRI 3.0 |
| This Work *“15sOp”* | 15 | 75.0% | 83.4% | 92.1% | NUIG18_17_C3 |
| Frenklach *DRM19* | 21 | 91.5 % | 94.3% | 96.2% | GRI 1.2 |
| Frenklach *DRM22* | 24 | 98.1% | 94.5% | 98.0% | GRI 1.2 |
| Lytras | 22 | 23.1% | 85.4% | 36.3% | USC Mech II |
| This Work *“19sOp”* | 19 | 93.6% | 94.9% | 95.6% | NUIG18_17_C3 |

Due to the relatively poorer performance of 15sOp at atmospheric conditions (74%), the overall flame speed fidelity of 15sOp is similar to that of the model created by Leylegian. However, 15sOp considerably outperforms the model of Leylegian’s at 0D conditions.

19sOp performs similarly or better than other models in literature that contain more species across the range of 0D and 1D targets.

## CONCLUSION

A novel three-step methodology to produce highly reduced and optimised “compact” kinetic models for methane combustion was proposed and demonstrated. The methodology emphasizes the capability of modern computing facilities to perform large numbers of simulations quickly. It relies on this capability to learn discreet combinations of numerical terms that will result in an accurate replication of complex combustion phenomena. Though much is known about the fundamentals of methane combustion, this methodology explicitly limits the use of this information for the purpose of advancing the development of fully automated machine learning protocols that can create minimally-sized, yet accurate, combustion kinetic models in an automated manner.

To describe methane combustion, two minimally detailed reaction networks of 19 and 15 species were first created. The "effective" reaction rate constants describing the chemical reactions of these models were then optimised using the TCD in-house MLOCK algorithm. In this study, only the Arrhenius *A*, *n* and *E* rate constant parameters were optimised. The first target of optimisation is the OH, H, CO and $CH_4$ profiles computed by the parent detailed model in PSR simulations. The second optimisation target is the laminar flame speed calculations of the detailed model. To find the optimum set of new Arrhenius parameters for each optimisation phase, coarse grid-like scans were performed to initially parametrize the rate constant bound space. From this, areas of the bound space which produce well performing models were identified and further resolved by higher resolution scans. From the scans a "genetic seed" was selected and used to direct the search algorithm to promising bound space regions to improve the efficiency and effectiveness of the entire procedure.

The best performing models created by this process are labelled "19sOp" and "15sOp". Each is then tested against a broader range of unseen gas turbine relevant performance targets.

It is found that both 15sOp and 19sOp models perform well against all 0D and 1D validation targets. 19sOp outperforms 15sOp as expected, due to the greater number of species and therefore greater level of information and degrees of freedom contained in this reaction network. Although the models perform best at operating conditions at which they were optimised, they also perform well at conditions outside the optimisation conditions. However, the performance decreases as the pressure is reduced to atmospheric. To improve the model's performance here, an atmospheric condition could be included in the objective error function which automatically selects the best performing models.

15sOp was shown to outperform the current state-of-art models in terms of 0D and 1D validation tests across the large range of conditions of interest. The larger 19sOp was also shown to outperform models of similar size from the literature, even those that contain more species. Therefore, if one can afford to use a reduced model with 20-25 species, 19sOp offers ~10% greater accuracy than 15sOp. Both models offer considerable efficiency savings relative to the use of the detailed model. 15sOp and 19sOp respectively result in a speed-up factor of 79 and 106, for fidelities of 93-98% and 75-91% relative to the detailed model (100%) in the simulation of a broad range of 0D and 1D methane combustion behaviours.

## ACKNOWLEDGEMENTS

The research reported in this publication was support by funding from Siemens Canada Limited and the Sustainable Energy and Fuel Efficiency (SEFE) Spoke of MaREI, the SFI Centre for Energy, Climate and Marine Research (16/SP/3829). Calculations were performed on the Boyle cluster maintained by the Trinity Centre for High Performance Computing.